\documentclass[final,3p,times]{elsarticle}

\usepackage{amssymb}

\usepackage{graphicx}
\usepackage[ansinew]{inputenc}

\journal{Microprocessors and Microsystems: Embedded Hardware Design}

\begin{document}

\begin{frontmatter}

\title{Simulation of High-Performance Memory Allocators}

\author[addr1]{Jos\'{e} L. Risco-Mart\'{i}n}
\author[addr2]{J. Manuel Colmenar}
\author[addr3]{David Atienza}
\author[addr1]{J. Ignacio Hidalgo}
\address[addr1]{Department of Computer Architecture and Automation, Universidad Complutense de Madrid, 28040 Madrid, Spain}
\address[addr2]{C. E. S. Felipe II, Universidad Complutense de Madrid, 28300 Aranjuez, Spain}
\address[addr3]{Embedded Systems Laboratory (ESL), Ecole Polytechnique F\'ed\'erale de Lausanne (EPFL), 1015 Lausanne, Switzerland}

\begin{abstract}
For the last thirty years, a large variety of memory allocators have been proposed. Since performance, memory usage and energy consumption of each memory allocator differs, software engineers often face difficult choices in selecting the most suitable approach for their applications. To this end, custom allocators are developed from scratch, which is a difficult and error-prone process. This issue has special impact in the field of portable consumer embedded systems, that must execute a limited amount of multimedia applications, demanding high performance and extensive memory usage at a low energy consumption. This paper presents a flexible and efficient simulator to study \textit{Dynamic Memory Managers (DMMs)}, a composition of one or more memory allocators. This novel approach allows programmers to simulate custom and general DMMs, which can be composed without incurring any additional runtime overhead or additional programming cost. We show that this infrastructure simplifies DMM construction, mainly because the target application does not need to be compiled every time a new DMM must be evaluated and because we propose a structured method to search and build DMMs in an object-oriented fashion. Within a search procedure, the system designer can choose the ``best'' allocator by simulation for a particular target application and embedded system. In our evaluation, we show that our scheme delivers better performance, less memory usage and less energy consumption than single memory allocators.
\end{abstract}


\end{frontmatter}

\section{Introduction and related work}
\label{sec:Introduction}

Current multimedia applications tend to make intensive use of dynamic memory due to their inherent data management. For this reason, many general-purpose memory allocators have been proposed to provide good runtime and low memory usage for a wide range of applications \cite{Johnstone1999, Lea}. However, using specialized memory allocators that take advantage of application-specific behavior can dramatically improve application performance \cite{Barrett1993, Grunwald1993, Wilson1995}. In fact, three out of the twelve integer benchmarks included in SPEC (parser, gcc, and vpr \cite{SPEC}) and several server applications, use one or more custom allocators \cite{Berger2001}.

Studies have shown that dynamic memory management can consume up to 38\% of the execution time in C++ applications \cite{Calder1995}. Thus, the performance of dynamic memory management can have a substantial effect on the overall performance of C++ applications. In this regard, programmers write their own ad hoc custom memory allocators as macros or monolithic functions in order to avoid function call overhead. This approach, implemented to improve application performance, is enshrined in the best practices of skilled computer programmers \cite{Meyers1995}. Nonetheless, this kind of code is brittle and hard to maintain or reuse, and as the application evolves, it can be difficult to adapt the memory allocator as the application requirements vary. Moreover, writing these memory allocators is both error-prone and difficult. Indeed custom and efficient memory allocators are complicated pieces of software that require a substantial engineering effort.

Therefore, to design ``optimal'' memory allocators, flexible and efficient infrastructures for building custom and general-purpose memory allocators have been presented in the last decade \cite{Berger2001, Atienza2006a, Atienza2006}. All the proposed methodologies are based on high-level programming where C++ templates and object-oriented programming techniques are used. It allows the software engineer to compose several general-purpose or custom memory allocator mechanisms. The aforementioned methodologies enable the implementation of custom allocators from their basic parts (e.g., de/allocation strategies, order within pools, splitting, coalescing, etc.). In addition, \cite{Atienza2006a} and \cite{Atienza2006} provide a way to evaluate the memory and energy used by the memory allocator, but at system-level (i.e., once the custom allocator has been designed). However, all the mentioned approaches require the execution of the target application to evaluate every candidate custom allocator, which is a very time-consuming task, especially if the target application requires human input (like video games). In this regard, \cite{Lo2004} and \cite{Teng2008} presented two memory manager design frameworks that allows the definition of multiple memory regions with different disciplines. However, these approaches are limited to a small set of user-defined functions for memory de/allocation.

Although most of the efforts on the optimization of the dynamic memory management have been focused on performance, there are two additional metrics that cannot be overlooked: memory usage and energy consumption. On the one hand, new embedded devices must rely on dynamic memory for a very significant part of their functionality due to the inherent unpredictability of the input data. These devices also integrate multiple services such as multimedia and wireless network communications which also compete for a reduced memory space. Then, the dynamic memory management influences the global memory usage of the system \cite{Atienza2006a}. On the other hand, energy consumption has become a real issue in overall system design (both embedded and general-purpose) due to circuit reliability and packaging costs \cite{Vijaykrishnan2003}. 

As a result, the optimization of the dynamic memory subsystem has three goals that cannot be seen independently: performance, memory usage and energy consumption. Unfortunately, due to these conflicting optimization metrics, there cannot exist a memory allocator that delivers the best performance and least memory and energy usage for all programs. However, a custom memory allocator that works best for a particular program can be developed \cite{RiscoMartin2009b}. Thus, new approaches to measure performance, memory usage and energy consumption are needed when designing a custom or general-purpose memory allocators.

In this paper we present a flexible, stable and highly-configurable simulator of memory allocators. By profiling of the target application, the proposed simulator can receive offline the dynamic behavior of the application and evaluate all the aforementioned metrics. As a result, the simulator can be integrated into a search mechanism in order to obtain optimum allocators.

The rest of the paper is organized as follows. First, Section \ref{sec:DynamicMemoryManagement} describes the design space of memory allocators. Then, Section \ref{sec:TheFramework} details the design and implementation of the simulation framework, as well as different configuration examples. Next, Section \ref{sec:SetUp} shows our experimental methodology, presenting the six benchmarks selected, whereas Section \ref{sec:Experiments} shows the results for these benchmarks. Finally, Section \ref{sec:Conclusions} draws conclusions and future work.


\section{Dynamic Memory Management}\label{sec:DynamicMemoryManagement}

In this Section, we summarize the main characteristics of dynamic memory management, as well as a classification of memory allocators, which we subsequently use in the implementation of the simulator.

\subsection{Dynamic Memory Management}

Dynamic memory management basically consists of two separate tasks, i.e., allocation and deallocation. Allocation is the mechanism that searches for a memory block big enough to satisfy the memory requirements of an object request in a given application. Deallocation is the mechanism that returns a freed memory block to the available memory of the system in order to be reused subsequently. In current applications, the blocks are requested and returned in any order. The amount of memory used by the memory allocator grows up when the memory storage space is used inefficiently, reducing the storage capacity. This phenomenon is called fragmentation. Internal fragmentation happens when requested objects are allocated in blocks whose size is bigger than the size of the object. External fragmentation occurs when no blocks are found for a given object request despite enough free memory is available. Hence, on top of memory de/allocation, the memory allocator has to take care of memory usage issues. To avoid these problems, some allocators include splitting (breaking large blocks into smaller ones to allocate a larger number of small objects) and coalescing (combining small blocks into bigger ones to allocate objects for which there are no available blocks of their size). However, these two algorithms usually reduce performance, as well as consume more energy. To support these mechanisms, additional data structures are built to keep track of the free and used blocks.

There exist several general-purpose allocators. Here we briefly describe two of them, the Kingsley allocator \cite{Wilson1995} and the Lea allocator \cite{Lea}. Kingsley and Lea are widely used in both general-purpose and embedded systems, and they are on opposite ends between maximizing performance and minimizing memory usage.

The Kingsley allocator was originally used in BSD 4.2, and current Windows-based systems (both mobile and desktop) apply the main ideas from Kingsley. The Kingsley allocator organizes the available memory in power-of-two block sizes: all allocation requests are rounded up to the next power of two. This rounding can lead to severe memory usage issues, because in the worst case, it allocates twice as much memory as requested. It performs no splitting (breaking large blocks into smaller ones) or coalescing (combining adjacent free blocks). This algorithm is well known to be among the fastest allocators although it is among the worst in terms of memory usage \cite{Berger2001}.

On the contrary, the Lea allocator is an approximate best-fit allocator that provides mainly low memory usage. Linux-based systems use as their basis the Lea allocator. It presents a different behavior based on the size of the requested memory. For example, small objects (less than 64 bytes) are allocated in free blocks using an exact-fit policy (one linked list of blocks for each multiple of 8 bytes). For medium-sized objects (less than 128Kb), the Lea allocator performs immediate coalescing and splitting in the previous lists and approximates best-fit. For large objects ($\geq$ 128Kb), it uses virtual memory (through \texttt{mmap}).

\subsection{Classification of memory allocators}

Memory allocators are typically categorized by the mechanisms that manage the lists of free blocks (free lists). These mechanisms include segregated free lists, simple segregated storage, segregated fit, exact segregated fit, strict segregated fit, and buddy systems.

\begin{figure}[!t]
\centering
\includegraphics[scale=1]{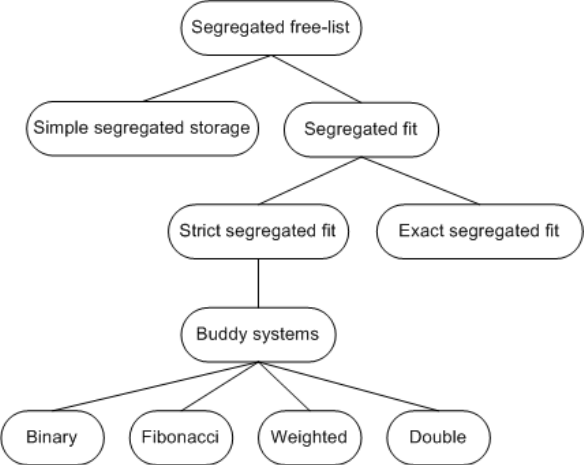}
\caption{Classification of memory allocators}
\label{fig:DmmClassification}
\end{figure}

\figurename~\ref{fig:DmmClassification} shows the classification of memory allocators provided in \cite{Wilson1995}. To easily define the design space in our simulator, we have reorganized these memory allocator mechanisms in a hierarchy. In the following we briefly describe these allocator mechanisms.

A segregated free-list allocator divides the free list into several subsets, according to the size of the free blocks. A freed or coalesced block is placed on the appropriate list. An allocation request is served from the appropriate list. This class of mechanism usually implements a good fit or best fit policy.

Simple segregated storage is a segregated free-list allocation mechanism which divides the storage into pages or other areas, and only allocates objects of a single size, or small range of sizes, within each area. This approach makes allocation fast and avoids headers, but may lead to high external fragmentation, as unused parts of areas cannot be reused for other object sizes.

Segregated fit is another variation of the segregated free-list class of allocation mechanisms. It maintains an array of free lists, each list holding free blocks of a particular range of sizes. The manager identifies the appropriate free list and allocates from it (often using a first-fit policy). If this mechanism fails, a larger block is taken from another list splitting it accordingly.

Exact segregated fit is a segregated fit allocator, which has a separate free list for each possible block size. The array of free lists may be represented sparsely. Large blocks may be treated separately. The details of the mechanism depend on the distribution of sizes between free lists.

Strict segregated fit is a segregated fit allocation mechanism which has only one block size on each free list. A requested block size is rounded up to the next provided size, and the first block on that list is returned. The sizes must be chosen so that any block of a larger size can be split into a number of smaller sized blocks.

Buddy systems are special cases of strict segregated fit allocators, which make splitting and coalescing fast by pairing each block with a unique adjacent buddy block. To this end, an array of free lists exists, namely, one for each allowable block size. Allocation rounds up the requested size to an allowable size and allocates from the corresponding free list. If the free list is empty, a larger block is selected and split. A block may only be split into a pair of buddies. A block may only be coalesced with its buddy, and this is only possible if the buddy has not been split into smaller blocks. Different sorts of buddy system are distinguished by the available block sizes and the method of splitting. They include binary buddies (the most common type), Fibonacci buddies, weighted buddies, and double buddies \cite{Wilson1995}.

\begin{table}[ht]
\caption{A DMM composed by two allocators.}
\begin{center}
\begin{tabular}{ccc}
\hline
\multicolumn{3}{c}{BuddySystemBinary, split=false, coalesce=false} \\
Data Structure & Mechanism(Policy) & Range (bytes) \\
DLL & FIRST(FIFO) & (0, 1] \\
DLL & FIRST(FIFO) & (1, 2] \\
DLL & FIRST(FIFO) & (2, 4] \\
DLL & FIRST(FIFO) & (4, 8] \\
\hline
\multicolumn{3}{c}{SegregatedFreeList, split=false, coalesce=true} \\
Data Structure & Mechanism(Policy) & Range (bytes) \\
SLL & FIRST(LIFO) & (8, 1490944] \\
\hline
\end{tabular}
\end{center}
\label{tab:DmmExample}
\end{table}

In the following, we define a \textit{Dynamic Memory Manager (DMM)} as a composition of one or more memory allocators. For instance, \tablename~\ref{tab:DmmExample} shows a DMM composed by two allocators. The first one is binary buddy system. It does not allow us splitting and/or coalescing. In addition, it manages block sizes in the range of $(0 \ldots 8]$ bytes. The data structure used to store blocks is a doubly-linked list. Finally, the allocation mechanism and allocation policy are first fit and first in first out, respectively. The second allocator follows a segregated free list behavior. Without spliting and/or coalescing, it manages block sizes in the range of $(8 \ldots 1490944]$ bytes. The data structure used to store blocks is a singly-linked list. Allocation mechanism and policy are first fit and last in first out, respectively.


\section{DMM Simulation Framework}\label{sec:TheFramework}

In this section we motivate and describe our proposed simulation approach as well as outline its design goals. In fact, as introduced in Section \ref{sec:Introduction}, there are currently several libraries to implement general-purpose and custom DMMs \cite{Lea, Berger2001, Atienza2006}. However, exploration techniques cannot be easily applied. Indeed, each custom design must be implemented, compiled and validated against a target application; hence, even if the DMM library is highly modular, this is a very time-consuming process. Thus, a simulator can greatly help in such optimization by being part of a higher optimization module that allows system designers to evaluate (in terms of performance, memory usage and energy consumption) the DMM for the target application. Therefore, the desired design goals for the development of this DMM exploration framework are:
\begin{itemize}
    \item Efficiency: since the simulator needs to be included into search algorithms, the DMM simulator must improve the execution time of a real DMM.
	\item Flexibility: software engineers must be able to simulate any DMM as a composition of single memory allocators. Thus, the parameters of each allocator should be highly configurable.
\end{itemize}

\begin{figure}[!t]
\centering
\includegraphics[scale=1]{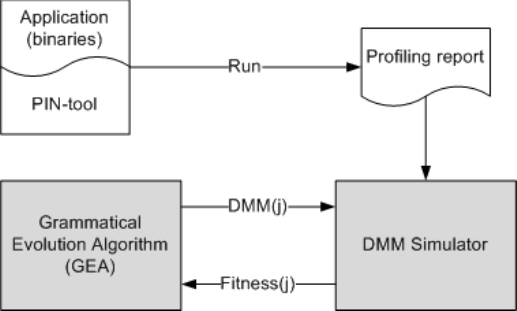}
\caption{DMM generation and evaluation process.}
\label{fig:Framework}
\end{figure}

\figurename~\ref{fig:Framework} shows an illustrative example on how our proposed methodology operates. In order to work with our simulator, the engineer must start with a profiling phase of the target application. To this end, we have used a tool called Pin \cite{Luk2005}. Pin's instrumentation is easy to work with because it allows the designer to insert calls to instrumentation at arbitrary locations in the target application source code. The Pin distribution includes many sample architecture-independent Pin-tools including profilers, cache simulators, trace analyzers, and memory bug checkers. Then, the profiling report is obtained using a modified version of the \texttt{malloctrace.cpp} example provided by Pin. As a result, after running the application, a profiling report is available and the system designer can test different DMMs using the same profiling report. Thus, the application must be executed just once during the whole study.

According to \figurename~\ref{fig:Framework}, the input to the simulator is a profiling report, which logs all the blocks that have been de/allocated. Our search algorithm is based on \textit{Grammatical Evolution} \cite{Brabazon2008} (details about its implementation can be found in \cite{RiscoMartin2009b}). This algorithm is constantly generating different DMM implementation candidates. Hence, when a DMM is generated ($DMM(j)$ in \figurename~\ref{fig:Framework}), it is received by the DMM simulator. Next, the DMM simulator emulates the behavior of the application, debugging every line in the profiling report. Such emulation does not de/allocate memory from the computer like the real application, but maintains useful information about how the structure of the selected DMM evolves in time. Then the profiling report is simulated and the simulator returns back the fitness of the current DMM to the search algorithm. After a given number of iterations, the search algorithm is able to find a custom DMM optimized for the target application in terms of performance, memory usage and energy consumption.

Our simulator framework can be mainly used in two different ways: (a) performing an automatic exploration as described in \figurename~\ref{fig:Framework}, or (b) selecting the best design among a predefined set of DMM candidates. Next, we describe the procedure to perfom such tests.

As a consequence of the simulator design, the composition of a DMM candidate follows a straightforward process for the designer. For instance, one common way of improving memory allocation performance is allocating all objects from a highly-used class from a per-class pool of memory. Because all these objects are of the same size, memory can be managed by a simple singly-linked free-list. Thus, programmers often implement these per-class allocators in C++ by overloading the \texttt{new} and \texttt{delete} operators for the class. This ad-hoc memory allocator can be easily implemented using an exact segregated fit allocation mechanism. We show next how we can use our library to compose such complex DMM.

\begin{verbatim}
ProfilingReport profReport = new ProfilingReport();
profReport.load("profile.mem");
ExactSegregatedFit exact = new ExactSegregatedFit(0
                                                  , profReport.getMaxSizeInB()
                                                  , profReport.getSizesInB());
exact.setup(FreeList.DATA_STRUCTURE.SLL
            , FreeList.ALLOCATION_MECHANISM.FIRST
            , FreeList.ALLOCATION_POLICY.FIFO);
DynamicMemoryManager manager = new DynamicMemoryManager();
manager.add(exact);
\end{verbatim}

In this example, when we build the \textit{ExactSegregatedFit} allocator, we provide the constructor with the minimum block size in bytes, the maximum block size in bytes and the different sizes supported by this allocator. In this example, the last two parameters are given by the profiling report, but they can be set manually, i.e., we can use the generated profiling report (as illustrated in the example), or we can study the different classes used in the application as well as their sizes and introduce them in the code above. After that, we configure the allocator defining the data structure to be used (singly-linked list) the allocation mechanism (first-fit) and the allocation policy (first in, first out). Finally, we build the corresponding DMM. As defined before, a DMM may contain of one or more allocators in our case.

\begin{figure}[ht]
\centering
\includegraphics[scale=0.418]{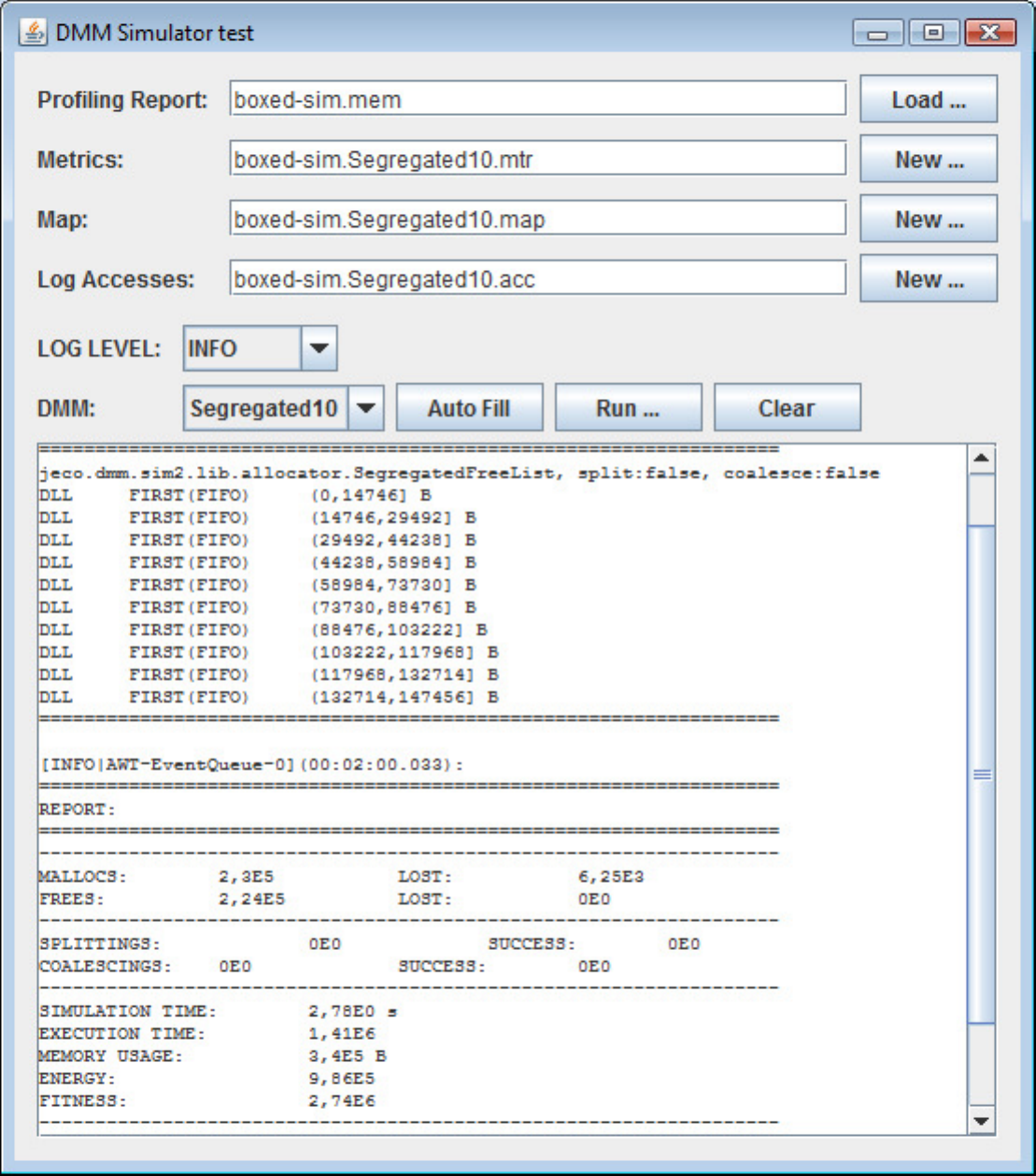}
\caption{Simulator Graphical User Interface}
\label{fig:SimulatorGui}
\end{figure}

Finally, the simulator is invoked, and after a few seconds we obtain all the metrics needed to evaluate the current DMM:

\begin{verbatim}
Simulator simulator = new Simulator(profReport, manager);
simulator.start();
simulator.join();
\end{verbatim}

However, as \figurename~\ref{fig:SimulatorGui} shows, to facilitate the use of the simulator, we have developed a GUI to test some general-purpose memory allocators, as well as to perform an automatic exploration of DMMs for the target application. This type of interface is very important for DMM exploration because, given a profiling report, the interface is able to simulate the selected allocator, giving its ``map'' and multiple metrics to the designer. Thus, we can dinamically control the exploration while it is performed. In any case, all the metrics are saved in external files for the designer be able to perform any desired post-processing analysis in order to explore alternative design space search options.

As a result, in our proposed exploration framework, at the same time that the simulation runs, several relevant metrics are computed, such as number of de/allocations, splittings, coalescings, performance, memory usage, memory accesses, etc. All the previous parameters except the execution time can be calculated accurately. However, since the system is using simulation time instead of real time, the total execution time is calculated as the computational complexity (or time complexity) \cite{Sipser1996}. Finally, the energy is computed using the execution time, memory usage and memory accesses, following the model described in \cite{RiscoMartin2009b}. The following code snippet shows an illustrative example of how the time complexity and memory accesses are calculated in the proposed DMM simulation framework:

\begin{verbatim}
void firstFit(long sizeInB) {
// ...
  while(iterator.hasNext()) {
    counterForMetrics++;
    currentBlock = iterator.next();
    if(currentBlock.sizeInB>=sizeInB) {
      block = currentBlock;
      iterator.remove();
      break;
    }
  }

  metrics.addExecutionTime(counterForMetrics);
  metrics.addMemoryAccesses(2*counterForMetrics);
  // ...
}
\end{verbatim}

The previous code excerpt shows a portion of the private function first-fit inside the simulator. The main loop looks for the first block big enough to allocate the requested size. We count the number of iterations in the loop, and after that, both the execution time and memory accesses are updated accordingly (+1 for each cycle in the loop to compute the computational time, and +2 for each cycle to count two accesses in the actual allocator, as one is needed to access to the current node in the free-list and another one to compute the size, i.e., subtraction of two pointers).

\begin{figure}[ht]
\centering
\includegraphics[scale=0.75]{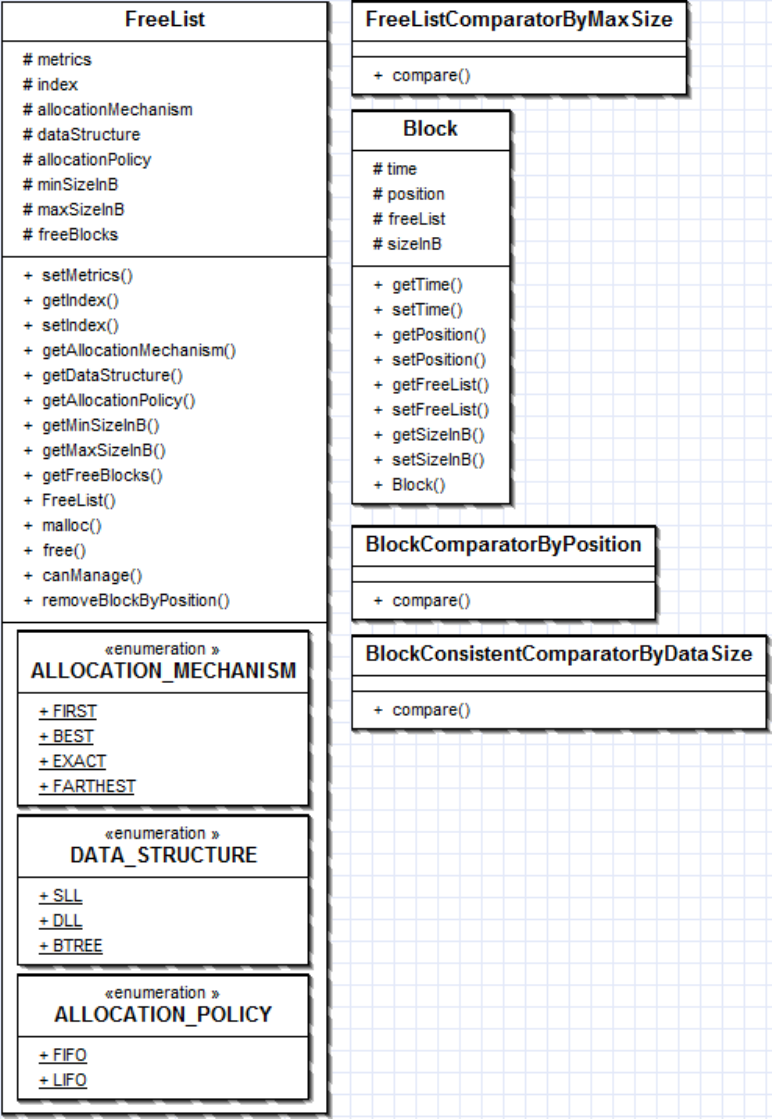}
\caption{Freelist package class diagram}
\label{fig:ClassDiagramFreelistPackage}
\end{figure}

The global design of the simulator follows an object-oriented architecture. Available at \cite{JECO}, it contains three main classes on the top layer: (1) a DMM class, which allows us to simulate standard allocation operations like \texttt{malloc()} or \texttt{free()}, (2) the Simulator class, which reads each line in the profiling report and calls the corresponding function in the DMM class, i.e., \texttt{malloc()} or \texttt{free()}, and (3) the Fitness class, which evaluates the DMM in terms of performance, memory usage and energy consumption. Our DMM simulation framework also includes Graphical User Interfaces (GUIs) to facilitate some standard simulations (based on general-purpose DMMs). The kernel of the DMM simulator includes all the allocation mechanisms described in Section \ref{sec:DynamicMemoryManagement}, as well as two general-purpose DMMs: Kingsley and Lea. Finally, some extra required functionality has been implemented: reading of the profiling report, analysis of the current block sizes and frequency of use, as well as number of mallocs, frees, invalid mallocs (mallocs without frees), invalid frees (frees without malloc), memory operations per second, reads, writes, etc.

The main simulation loop, once the custom DMM has been built with one or more allocators, works as follows: As mentioned above, every entry (malloc or free) in the profiling report is read by the Simulator class. For each line, the corresponding function in the DMM class is invoked. The DMM class contains a pool of allocated blocks. If the operation type is \textit{free}, the DMM class gets the block from that pool and inserts it into the corresponding allocator. On the contrary, if the operation type is \textit{malloc}, the DMM tries to extract a free block from the corresponding allocator. Thus, independently of the operation type, the DMM class tries to find the allocator responsible for managing the current block size. Again, the allocator tries to find the free-list in charge of the current block size (one allocator can manage one or more free-lists). Once the free-list has been found, the new block is inserted (free) or extracted (malloc). In the last case, if no free block is found, we use the virtual memory (the aforementioned pool), and external fragmentation occurs. However, if splitting or coalescing are allowed in the corresponding allocator, it tries to find the way to split a free block or to coalesce two adjancent blocks. 

Both allocators and free-lists are contained in a two-level hash table, where the key is the size interval that an allocator (or free-list, respectively) is able to manage. Obviously, if an allocator is for example a Binary Buddy System, all its free-lists are contained in an array, an thus, the access to a free-lists is performed in constant time O(1). As described above, the internal functionality of an allocator is quite simple, except in the case of splitting or coalescing, i.e., it just search for the corresponding free-list. The behavior implementation of free-lists is more complex. To better understand the internal functionality, we present \figurename~\ref{fig:ClassDiagramFreelistPackage}. In this \figurename, first, a free-list contains a set of free-blocks (\textit{freeBlocks} attribute in \figurename~\ref{fig:ClassDiagramFreelistPackage}, class FreeList). The container responsible for managing these blocks can be a singly-linked list, a doubly-linked list or a binary tree (see DATA\_STRUCTURE in the FreeList class). Every block in the free-list is instanciated from the Block class. It contains information like time of creation, position (equivalent to memory address), a pointer to the free-list that contains the block, and the block size in bytes (size includes headers of the free-list). Blocks are extracted from the free-list following one of the four allocation mechanisms implemented (see ALLOCATION\_MECHANISM in \figurename~\ref{fig:ClassDiagramFreelistPackage}, class FreeList): first fit, best fit, exact fit and farthest fit (farthest in memory address from the hottest block, used for temperature-aware analysis in \cite{Colmenar2010}). Blocks are also extracted using one of the two implemented allocation policies: first in first out or last in first out (see ALLOCATION\_POLICY in \figurename~\ref{fig:ClassDiagramFreelistPackage}, class FreeList). Hence, the object-oriented structure proposed in this framework, enables the final user to easily implement new allocator mechanisms and policies.


\section{Experimental Methodology}\label{sec:SetUp}

In this section, we study the performance, memory usage and energy consumption implications of building general-purpose and custom allocators using the simulator. To evaluate those three metrics, we have selected a number of memory-intensive programs, listed in \tablename~\ref{tab:Benchmarks}, namely:

\begin{table*}[ht]
	\caption{Statistics for the analyzed memory-intensive benchmarks}
	\centering
	\begin{tabular}{llllll}
\hline
\multicolumn{6}{c}{\textbf{Memory-Intensive Benchmark Statistics}} \\
\hline
Benchmark     & Objects & Total memory (bytes) & Max in use (bytes) & Average size (bytes) & Memory ops \\
\hline
lindsay.mem   & 102143	& 5795534    & 1497149  & 56.74   & 204153 \\
boxed-sim.mem & 229983	& 2576780    & 339072   & 11.20	  & 453822 \\
cfrac.mem     & 570014	& 2415228    & 6656     & 4.24	  & 1140009 \\
gcbench.mem   & 843969	& 2003382000 & 32800952 & 2373.76 & 1687938 \\
espresso.mem  & 4395491	& 2078856900 & 430752   & 472.95  & 8790549 \\
roboop.mem    & 9268234	& 321334686  & 12802    & 34.67   & 18536420 \\
\hline
	\end{tabular}
	\label{tab:Benchmarks}
\end{table*}

- \textit{Lindsay} is a hypercube network communication simulator coded in C++ \cite{Berger2001}. It represents a hypercube network, then simulates random messaging, while accumulating statistics about the messaging performance. The hypercube is represented as a large array, which is alive for the entire run, while the messages are represented by small heap-allocated objects, which live very briefly, just long enough for the message to reach its destination.

- \textit{Boxed-sim} is a graphics application that simulates spheres bouncing in a box \cite{Chilimbi2001}. The application represents a physical experiment where gravity is turned off; thus, there is zero friction and no energy loss at each collision. Consequently, each sphere is given a random initial position, orientation and velocity, and zero initial angular velocity. Each run simulates a given amount of virtual time. 

- \textit{Cfrac} performs the factorization of an integer to two nearly equal factors \cite{Zorn1993}. It applies the continued fraction factorization algorithm, which is one of the fastest prime factorization algorithms. The benchmark uses dynamic memory by storing data through small allocations mostly between 2 and 40 bytes.

- \textit{GCBench} is an artificial garbage collector benchmark that allocates and drops complete binary trees of various sizes [17]. It maintains some permanent live data structures and reports time information for those operations. This benchmark appears to have been used by a number of vendors to aid in the Java VM development. That probably makes it less desirable as a means to compare VMs. It also has some know deficiencies, e.g. the allocation pattern is too regular, and it leaves too few ``holes'' between live objects. It has been recently proposed as the most useful sanity test to be used by garbage collector developers.

- \textit{Espresso} is an optimization algorithm for PLAs that minimizes boolean functions \cite{SPEC}. It takes as input a boolean function and produces a logically equivalent function, possibly with fewer terms. Both the input and output functions are represented as truth tables. The benchmark performs set operations, such as union, intersect and difference. In particular, the sets are implemented as arrays of unsigned integers and set membership is indicated by a particular bit being on or off. These data structures are instantiated using dynamic memory.

- \textit{Roboop} is a C++ robotics object-oriented programming toolbox suitable for synthesis and simulation of robotic manipulator models in an environment that provides ``MATLAB like'' features for the treatment of matrices \cite{Berger2001}. The test computes several robotics operations like forward and reverse kinematic, jacobian, torque and acceleration, involving dynamic memory operations for matrices and robotic components.

To perform the profiling of the applications, we applied the Pin tool, as described in Section \ref{sec:TheFramework}, and we run all the applications on an Intel Core 2 Quad processor Q8300 system with 4 GB of RAM, under Windows 7. This task was performed just once within the proposed DMM simulation architecture.

\tablename~\ref{tab:Benchmarks} includes the number of objects allocated and their average size. The applications' memory usages range from just 6.5 KB (for Cfrac) to over 31.28 MB (for GCBench). For all the programs, the ratio between total allocated memory and the maximum amount of memory in use is large. In addition, the number of memory operations is also large. Thus, all the proposed benchmarks highly rely on the dynamic memory subsystem. In the next step we simulate five general-purpose allocators, as well as performing an automatic exploration of feasible DMMs using a search algorithm (i.e., grammatical evolution in our case \cite{Colmenar2010}).


\section{Results}\label{sec:Experiments}

We simulated the benchmarks described in the previous section by considering five general-purpose allocators: the Kingsley allocator (labeled as KNG in the following figures), the Doug's Lea allocator (labeled as LEA), a buddy system based on the Fibonacci allocation algorithm (labeled as FIB), a list of 10 segregated free-lists (S10), and an exact segregated free list allocator (EXA). Finally, we compared the results with the custom DMM obtained with our proposed automatic exploration process (labeled as GEA).

To create the custom DMMs, we have followed the proposed methodology flow (\figurename~\ref{fig:Framework}). We first profiled the behavior of the application under study using the Pin tool. Next, we execute our grammatical evolutionary algorithm. To find a solution to our multiobjective optimization problem, we construct a single aggregate objective function.

\begin{figure}[ht]
\centering
\includegraphics[width=0.618\columnwidth]{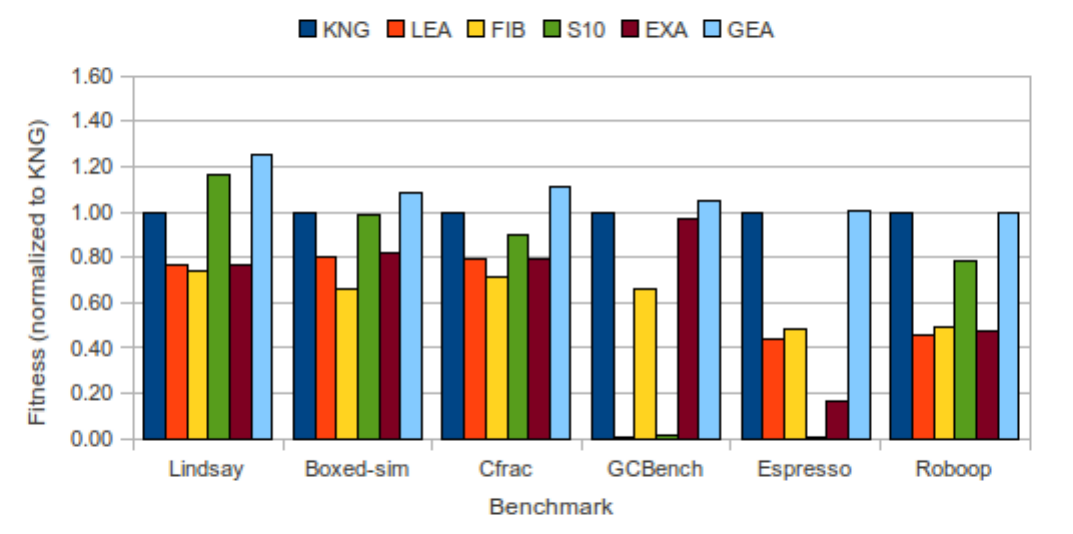}
\caption{Global fitness for the six benchmarks managed by six DMMs, normalized to the Kingsley allocator (in the sense that greater than 1 is better)}
\label{fig:ResultsFitness}
\end{figure}

\begin{figure}[ht]
\centering
\includegraphics[width=0.618\columnwidth]{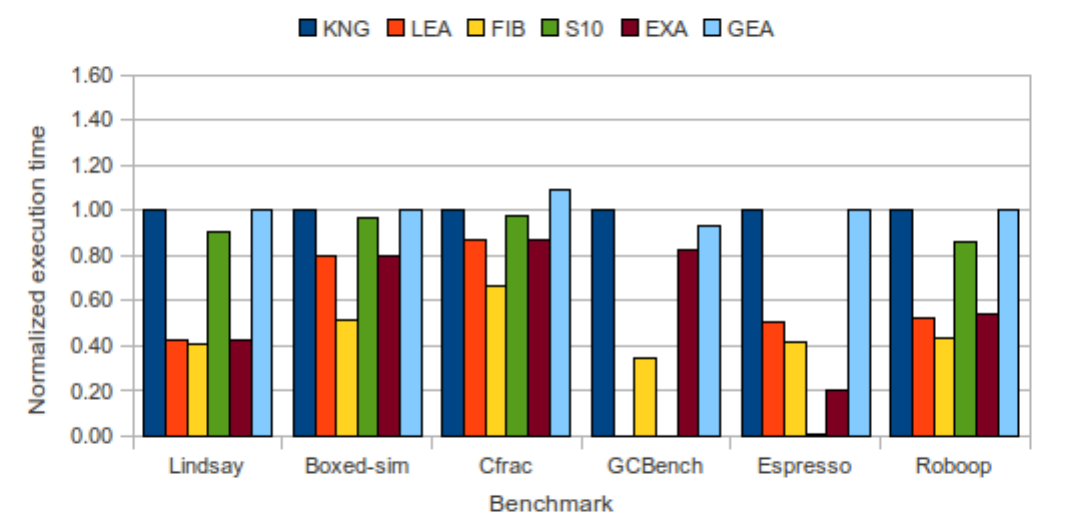}
\caption{Execution time for the six benchmarks managed by six DMMs, normalized to the Kingsley allocator (greater than 1 is better)}
\label{fig:ResultsPerformance}
\end{figure}

\begin{figure}[ht]
\centering
\includegraphics[width=0.618\columnwidth]{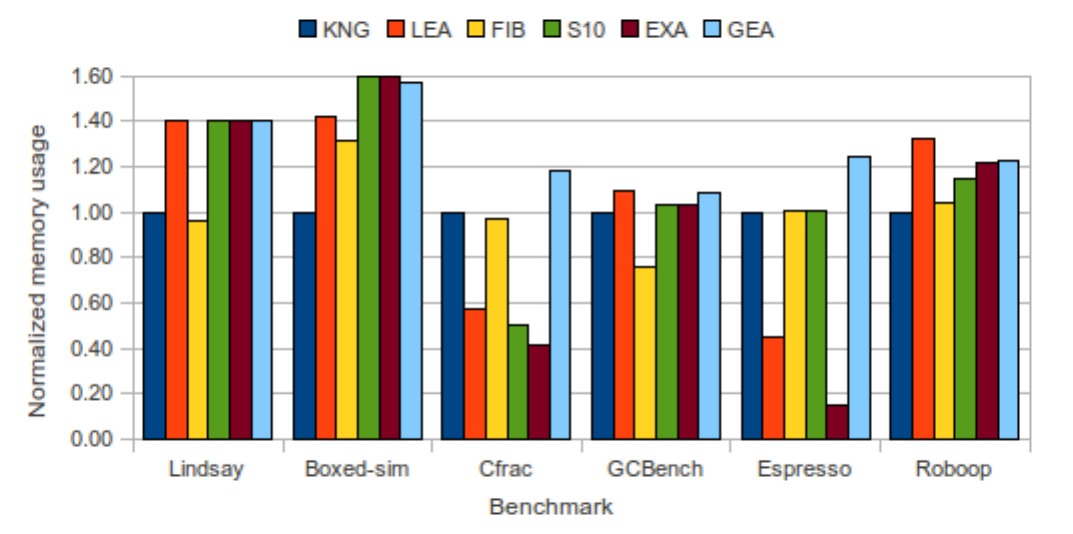}
\caption{Memory usage for the six benchmarks managed by six DMMs, normalized to the Kingsley allocator (greater than 1 is better)}
\label{fig:ResultsMemory}
\end{figure}

\begin{figure}[ht]
\centering
\includegraphics[width=0.618\columnwidth]{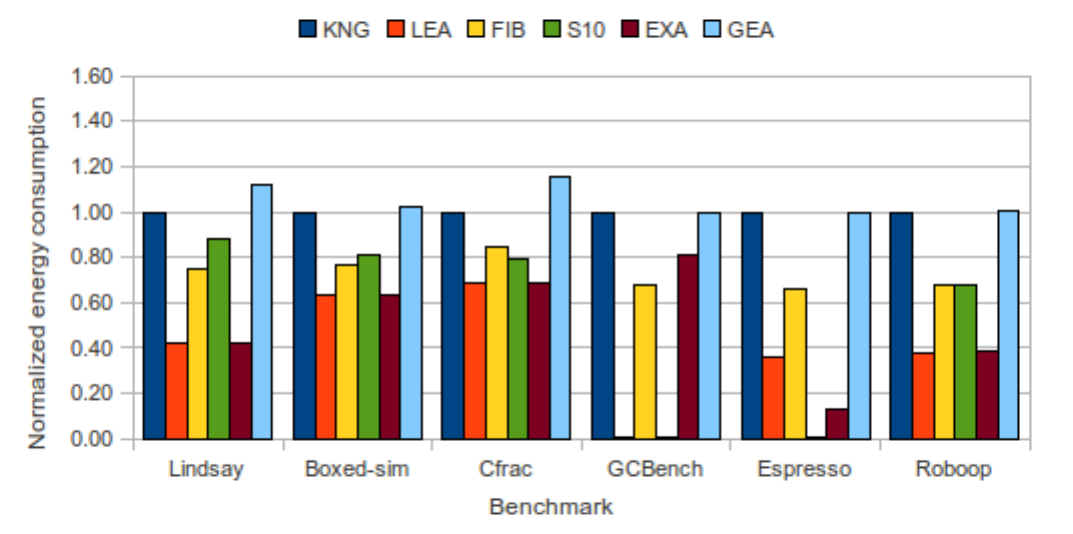}
\caption{Energy consumption for the six benchmarks managed by six DMMs, normalized to the Kingsley allocator (greater than 1 is better)}
\label{fig:ResultsEnergy}
\end{figure}

We present our experimental results in Figures \ref{fig:ResultsFitness}, \ref{fig:ResultsPerformance}, \ref{fig:ResultsMemory} and \ref{fig:ResultsEnergy}. To better scale the values, we have divided every Kingsley's metric by the corresponding one of all the DMMs in study. \figurename~\ref{fig:ResultsFitness} depicts the global fitness reached by all the proposed DMMs. As mentioned above, the global fitness is a weighted sum of execution time, memory used and energy consumed by the DMM. Figures \ref{fig:ResultsPerformance}, \ref{fig:ResultsMemory}, \ref{fig:ResultsEnergy} show each one of the components of the global fitness, i.e., execution time, memory usage and energy consumption respectively. Note that we define memory usage as the high water-mark of memory requested from the virtual memory.

As a global analysis of general-purpose DMMs, we can see in \figurename~\ref{fig:ResultsFitness} that Kingsley is an excellent allocator. It is the best general-purpose DMM not only in terms of performance (see \figurename~\ref{fig:ResultsPerformance}), but also considering energy consumed (see \figurename~\ref{fig:ResultsEnergy}). Only the memory usage (see \figurename~\ref{fig:ResultsMemory}) is the feature where KNG is outperformed, mainly by LEA, in four of six cases. As a consequence, LEA is a good candidate in order to reduce memory usage. The buddy system based on Fibonacci (FIB) always shows an intermediate behavior in all the six benchmarks. The other two general-purpose DMMs, labeled S10 and EXA, present a different behavior, depending on the benchmark. For example, examining the global fitness (\figurename~\ref{fig:ResultsFitness}), the ten segregated lists (S10) is a good allocator for Lindsay and Boxed-sim, but it is the worst choice for GCBench and Espresso. Finally, the exact segregated list (EXA) responds well for GCBench.

We now analyze the results obtained by GEA, examining each metric independently. To this end we also show in \tablename~\ref{tab:Percentage} the percentage of improvement reached by our exploration algorithm, compared to the Kingsley allocator. After a first view to the global fitness (see also \figurename~\ref{fig:ResultsFitness}) we may conclude that our custom DMM (labeled as GEA) is the best choice in all the six benchmarks, giving improvements of 20.46\%, 7.95\%, 10.01\%, 5.02\%, 0.57\% and 0.12\% in the cases of Lindsay, Boxed-sim, CFrac, GCBench, Espresso and Roboop, respectively.

\begin{table*}[ht]
\caption{Percentage of improvent. GEA vs. Kingsley}
\centering
\begin{tabular}{ccccccc}
\hline
 & Lindsay & Boxed-sim & CFrac & GCBench & Espresso & Roboop \\
\hline
Gobal Fitness   & 20.46\% & 7.95\% & 10.01\% & 5.02\% & 0.57\% & 0.12\% \\
Performance & 0.00\% & 0.00\% & 8.19\% & -7.36\% & 0.00\% & 0.00\% \\
Memory usage & 28.77\% & 36.20\% & 15.45\% & 7.73\% & 19.51\% & 18.67\% \\
Energy consumption & 11.13\% & 2.14\% & 13.30\% & -0.63\% & 0.00\% & 0.33\% \\
\hline
\end{tabular}
\label{tab:Percentage}
\end{table*}

Regarding performance (\figurename~\ref{fig:ResultsPerformance}) the best results in this case are obtained by Kingsley and GEA. Note that in the case of GCBench, GEA is 7.36\% worst than Kingsley. It is not relevant, because Kingsley is a highly optimized  allocator for performance, while tends to perform worse than other managers regarding to memory footprint and energy consumption \cite{Atienza2006a}. Then, Lea, the Fibonacci-based buddy allocator, Segregated list and Exact fit, on the contrary, have different behaviors depending on the application (particularly S10 and EXA). Thus, these four allocators are worse than Kingsley and GEA in all conditions.

In terms of memory usage (\figurename~\ref{fig:ResultsMemory}), we can observe that Kingsley is not the best allocator, as it suffers from a high level of internal fragmentation which in turn results in a larger memory utilization. Although Lea is highly optimized in memory usage, we can see that it does not perform well in the case of Cfrac and Espresso. This occurs because these benchmarks allocate a big amount of memory at the beginning of the application, and when the free blocks can be reused, they are too big for the new requests. This issue induces both internal and external fragmentation. Once more, the DMM obtained by GEA is the best choice: 28.77\%, 36.20\%, 15.45\%, 7.73\%, 19.51\%, 18.67\% and 58\% better than Kingsley in Lindsay, Boxed-sim, CFrac, GCBench, Espresso and Roboop, respectively.

Since energy depends on the execution time, in some cases the map of the energy consumption is quite equivalent to the execution time as shown by Lindsay and GCBench in Figs. \ref{fig:ResultsPerformance} and \ref{fig:ResultsEnergy}. However, energy consumed may vary greatly depending on the used data structure (singly-linked list, doubly-linked list, binary search tree, etc.), because it heavily influences in the number of memory accesses and thus, in the energy. However, it is not a necessary condition, so the energy consumption is an independent metric and must be studied separately. In this case, both Kingsley and GEA are good candidates for all the three benchmark problems. However, in Lindsay and GCBench, the DMM obtained by GEA requires significant less energy than the Kingsley allocator (11.13\% and 13.30\% less, respectively). As a result, GEA is the best choice.

As a conclusion, we can state the exact fit allocation, a common practice among object-oriented applications, is not the best choice, especially for Espresso and Roboop benchmarks. We can also observe that using our proposed DMM simulator in the memory exploration process enables the study of six different allocation mechanisms starting from a unique previous run of each target application. This point, that comes from joining the profiling task and the simulator features, enables large savings in the dynamic memory optimization exploration, providing great benefit to designers.

\begin{table}[ht]
\caption{Custom DMM map for the CFrac benchmark.}
\begin{center}
\begin{tabular}{ccc}
\hline
\multicolumn{3}{c}{BuddySystemBinary, split=true, coalesce=true} \\
Data Structure & Mechanism(Policy) & Range (bytes) \\
BTREE & FIRST(FIFO) & (0,1] \\
BTREE & FIRST(FIFO) & (1,2] \\
BTREE & FIRST(FIFO) & (2,4] \\
BTREE & FIRST(FIFO) & (4,8] \\
BTREE & FIRST(FIFO) & (8,16] \\
BTREE & FIRST(FIFO) & (16,32] \\
BTREE & FIRST(FIFO) & (32,64] \\
\hline
\multicolumn{3}{c}{SimpleSegregatedStorage, split=false, coalesce=false} \\
Data Structure & Mechanism(Policy) & Range (bytes) \\
BTREE & EXACT(LIFO) & (64,1724] \\
\hline
\multicolumn{3}{c}{BuddySystemBinary, split=false, coalesce=false} \\
Data Structure & Mechanism(Policy) & Range (bytes) \\
SLL & EXACT(LIFO) & (1724,2048] \\
SLL & EXACT(LIFO) & (2048,4096] \\
\hline
\end{tabular}
\end{center}
\label{tab:Cfrac}
\end{table}

Finally, we present the custom DMM obtained by our search algorithm for the CFrac benchmark in \tablename~\ref{tab:Cfrac}. CFrac uses 22 different block sizes, varying from 2 to 3616 bytes (3.53 KB).The automatically-obtained custom DMM presents 3 different internal allocators. The first one is a binary buddy, allowing the memory splitting and coalescing. All the seven free-lists are implemented as binary trees with a first fit allocation and FIFO policy. This first allocator manages block sizes in the range from 0 to 64 bytes. The second allocator follows a simple segregated storage mechanism and contains just one free-list, implemented as a binary tree with an exact-fit mechanism and LIFO policy. The binary tree has more sense in segregated lists because they may build blocks of different sizes. However, we let GEA to select BTREE in binary buddies in order to simplify the search space. The last allocator is again a binary buddy with two lists. It manages blocks varying from 1.68 to 4 KB, implemented with singly-linked lists and using an exact-fit mechanism and LIFO policy. In this case the most requested block sizes are 2, 4 and 40. Thus implementing the first region as a binary buddy with splitting and coalescing seems to be a good choice to minimize execution time and saves energy, whereas the segregated list improves memory usage.


\section{Conclusions and future work}\label{sec:Conclusions}

Dynamic memory management continues to be a critical part of many recent applications in embedded systems for which performance, memory usage and energy consumption is crucial. Programmers, in an effort to avoid the overhead of general-purpose allocation algorithms, write their own custom allocation implementations trying to achieve (mainly) a better performance level. Because both general-purpose and custom allocators are monolithic designs, very little code reuse occurs between allocator implementations. In fact, memory allocator are hard to maintain and, as a certain application evolves, it becomes very complex to adapt the memory allocator to the changing needs of each application. In addition, writing custom memory allocators is both error-prone and difficult. Overall, efficient multi-objective (performance, memory footprint and energy efficient) memory allocators are complicated pieces of software that require a substantial engineering and maintaining effort.

In this paper, we have described a simulation framework in which custom and general-purpose allocators can be effectively constructed and evaluated. Our framework allows system designers to rapidly build and simulate high-performance allocators, both general and custom ones, while overcoming the tedious task of multiple profiling steps for each allocator instance within a target application. In particular, the proposed methodology avoids the compilation and execution of the target application in a case-by-case basis. Thus, the design of a custom DMM is quickly and free of (initial) implementation errors.

In addition to the proposed one-time DMM simulation approach, we have developed a complete search procedure to automatically find optimized custom DMMs for the application in study. Using this methodology, we have designed custom DMMs for six different benchmark applications, which enable important energy and memory footprint savings with respect to state-of-the-art memory allocation solutions.


\section*{Acknowledgments}
This work has been supported by Spanish Government grants TIN2008-00508 and MEC Consolider Ingenio CSD00C-07-20811 of the Spanish Council of Science and Innovation. This research has been also supported by the Swiss NSF Research Project (Division II) Grant number 200021-127282.

\bibliographystyle{elsarticle-num}
\bibliography{bibliography}

\end{document}